\documentstyle[sprocl]{article}

\input{psfig.sty}

\def\Journal#1#2#3#4{{#1} {\bf #2}, #3 (#4)}

\def\PLB{{\em Phys. Lett.}  B}
\def\PRL{\em Phys. Rev. Lett.}
\def\PRD{{\em Phys. Rev.} D}

\def\PROG{\em Prog. Theo. Phys.}
\def\EPJC{{\em Eur. Phys. J.} C}
\def\IJMPA{{\em Int. J. Mod. Phys.} A}

\def\be{\begin{equation}}
\def\ee{\end{equation}}
\def\bea{\begin{eqnarray}}
\def\eea{\end{eqnarray}}

\begin{document}

\title{Is Cabibbo-Kobayasi-Maskawa Matrix Unitary?}

\author{C. S. Kim\footnote{kim@kimcs.yonseiac.kr, 
~~http://phya.yonsei.ac.kr/\~{}cskim/}}

\address{Dept. of Physics and IPAP, Yonsei Univ, Seoul 120-749, Korea} 

\author{H. Yamamoto\footnote{hitoshi@phys.hawaii.edu, 
~~http://www.phys.hawaii.edu/\~{}hitoshi/}}

\address{Dept. of Physics \& Atrophysics, Univ of Hawaii, HI 96822, USA}


\maketitle\abstracts{ 
First, we give summary of the present values of CKM matrix elements.
Then, we discuss whether CKM matrix is unitary or not,
and how we can find out if it is not unitary.
}

\section{Introduction}
The Cabibbo-Kobayashi-Maskawa (CKM) matrix \cite{CKM} in three generation is
\begin{eqnarray}
V_{\rm CKM} &=& \left(
   \begin{array}{ccc}
     V_{ud}   &   V_{us}   & V_{ub}  \\
     V_{cd}   &   V_{cs}   & V_{cb}  \\
     V_{td}   &   V_{ts}   & V_{tb}
   \end{array}
   \right) \\
&=& \left(
\begin{array}{ccc}
 c_{12}c_{13}       &  s_{12}c_{13}     &  s_{13}e^{-i\delta_{13}}  \\
-s_{12}c_{23}-c_{12}s_{23}s_{13}e^{i\delta_{13}} &
        c_{12}c_{23}-s_{12}s_{23}s_{13}e^{i\delta_{13}} & s_{23}c_{13}  \\
 s_{12}s_{23}-c_{12}c_{23}s_{13}e^{i\delta_{13}} &
       -c_{12}s_{23}-s_{12}c_{23}s_{13}e^{i\delta_{13}} & c_{23}c_{13} 
\end{array}
\right). \nonumber
\end{eqnarray}
Directly measured values \cite{PDG} of the elements of $2 \times 2$ Cabibbo
matrix, $V_{\rm C}$,  are
\begin{equation}
V_{\rm C} = \left(
   \begin{array}{cc}
     V_{ud}   &   V_{us}   \\
     V_{cd}   &   V_{cs}   
   \end{array}
   \right) =
\left(
   \begin{array}{cc}
     0.9740 \pm 0.0010  & 0.2196 \pm 0.0023  \\
     0.224 \pm 0.016    & 1.04 \pm 0.16
   \end{array}
   \right) .
\end{equation}
Only after assuming 3 generation unitarity, $V_{\rm C}$ becomes well known,
and can be parametrized with one parameter, 
$\lambda = \sin\theta_c  \approx 0.22$, within $90\%$ CL 
\begin{equation}
V_{\rm C} =
\left(
   \begin{array}{cc}
     0.9745 \sim 0.0010  & 0.217 \sim 0.224  \\
     0.217  \sim 0.224   & 0.9737 \sim 0.9753
   \end{array}
   \right) \simeq
\left(
   \begin{array}{cc}
     1 - \frac{\lambda^2}{2}  & \lambda \\
     -\lambda   & 1 - \frac{\lambda^2}{2}
   \end{array}
   \right) .
\end{equation}
We note that the directly measured values (2) still have relatively large
uncertainties of $\sim 5 \%$. 
Extension to (unitarized) three generation by Kobayashi and Maskawa leads
one non-trivial phase angle $\delta_{\rm KM}$. 
In Wolfenstein parametrization \cite{Wolf}, it can be approximated as 
\begin{equation}
V_{\rm KM} = \left(
   \begin{array}{ccc}
        &   & V_{ub}  \\
        &   & V_{cb}  \\
     V_{td} & V_{ts}  & V_{tb} 
   \end{array}
   \right) \approx
\left(
   \begin{array}{ccc}
       &   &  A \lambda^3 (\rho - i \eta) \\
       &   &  A \lambda^2  \\
    A \lambda^3 (1- \rho - i \eta)  &  -A \lambda^2  &  1
   \end{array}
   \right) ,
\end{equation}
where $\rho + i \eta \approx \sqrt{\rho^2 + \eta^2} \exp (i \delta_{13})$, and
we asumed $V_{tb}=1$.
In coming discussions, we first assume $V_{\rm CKM}$ being unitary, but later
we will investigate  possible non-unitarity of $V_{\rm CKM}$.

\section{Theoretical Determination of Elements of $|V_{\rm CKM}|$} 
\subsection{$|V_{cb}| = A \lambda^2$}

From the exclusive $B \to D^* l \nu$ decay,
\begin{equation}
\frac{d {\cal BR}}{d w}(B \to D^* l \nu) = \tau_B \frac{d \Gamma}{d w} = 
\tau_B \times [....] \times \sqrt{w^2 -1}|V_{cb}|^2 {\cal F}^2(w),
\end{equation}
where \cite{Neubert}
\begin{equation}
{\cal F}(w) = {\cal F}(1) (1 - \hat{\rho}^2(w-1) + ...),~~~~
{\cal F}(1)=0.924 \pm 0.027,
\nonumber 
\end{equation}
with $w \equiv v_B \cdot v_{D^*}$,
LEP and CLEO \cite{CLEO} measured 
$\frac{d {\cal BR}}{d w}|_{w \to 1},\hat{\rho}^2,\tau_B$ to obtain the value
$$
|V_{cb}|(B \to D^* l \nu)=0.0387 \pm 0.0031,~~~({\rm i.e.}~A\sim 0.8).
$$
Inclusive measurement of semileptonic total decay rate, 
$\Gamma_{s.l.}(B \to X_c l \nu)$, can also give \cite{Vqb}
the value $|V_{qb}|$
from \begin{eqnarray}
\Gamma_{s.l.}(B &\to& X_q l \nu) = \gamma_q^{\rm th} \times |V_{qb}|^2 \\ 
&\equiv& \left(\frac{G_F^2 m_b^5}{192 \pi^3}\right)  
\left[z_0\left(1 - \frac{2 \alpha_s}{3 \pi}g(\epsilon_q)\right)
+\frac{1}{2} z_0(G_b-K_b)-2z_1G_b+. \right] |V_{qb}|^2, \nonumber
\end{eqnarray}
where
\begin{eqnarray}
g(\epsilon)&=&(\pi^2-\frac{31}{4})(1-\epsilon)^2+\frac{3}{2}, \nonumber\\
z_0(\epsilon)&=&1-8\epsilon^2+8\epsilon^6-\epsilon^8-24\epsilon^4\ln\epsilon,~~~
 z_1(\epsilon)=(1-\epsilon^2)^4, \nonumber\\
K_b &=& -\lambda_1/m_b^2,~~~G_b=3\lambda_2/m_b^2,~~~
 \alpha_s=\alpha_s(m_b^2), \nonumber
\end{eqnarray}
with $\epsilon_q=\frac{m_q}{m_b},~-\lambda_1=\mu_\pi^2=(0.1\sim0.6)~{\rm
GeV}^2,~ \lambda_2=\frac{1}{4}(m_{B^*}^2-m_B^2)=0.12~{\rm GeV}^2$.
Now from the total semileptonic decay width, we can measure
$$
|V_{cb}|=\left[\frac{\Gamma_{s.l.}(B\to X_c l \nu)_{\rm exp}}
 {\Gamma_{tot}(B)_{\rm exp} \tau_B^{\rm exp} \gamma_c^{\rm th}}\right]^{1/2}=
 0.0419 \sqrt{\frac{{\cal BR}(B \to X_c l \nu)}{0.105}}
 \sqrt{\frac{1.55}{\tau_B}}(1 \pm 0.04)~.
$$
We remark that the uncertainty in determination of $\lambda_1=-\mu_\pi^2$,
the average kinetic energy of $b$-quark in $B$ meson,
is still large.

\subsection{$|V_{ub}|=A \lambda^3 \sqrt{\rho^2+\eta^2}$}

This is probably the most important element, and at the same time one of the
most difficult to be measured. Exclusively, we use the semileptonic decays,
$B\to \rho l \nu, \to \pi l \nu, ...$, which invoke large theoretical
uncertainties from hadronic form factors, and  their model dependences.
Recently by using data of large $q^2$ ($14 < q^2 < 21$ GeV$^2$) and 
large $E_l>2.3$ GeV in $B\to \rho \l \nu$, CLEO derived \cite{CLEO2}
$$|V_{ub}|=(3.23 \pm 0.24 \pm 0.25 \pm 0.58) \times 10^{-3},$$
where the last error is from model dependence.

As we can easily see from Eq. (7), if we  also measure the total decay width of
$\Gamma_{s.l.}(B \to X_u l \nu)$, then we can extract \cite{Vub}
$|V_{ub}|$ from
$$
\left|\frac{V_{ub}}{V_{cb}}\right| \simeq (0.81 \pm 0.06)\times
 \left[\frac{{\cal BR}(B \to X_u l \nu)}
        {{\cal BR}(B \to X_c l \nu)}\right]^{1/2},
$$
where the error is from $m_b,\alpha_s,\mu_\pi^2$.
However, the separation of $B \to X_u$ from the dominant $B\to X_c$ is
experimentally difficult. The promissing
method is to use $M_X$, hadronic invariant recoiled mass.
This is because \cite{Barger}
$$m_u \ll m_c~~~ \Longrightarrow~~~ m_{\pi,\rho} \ll m_{D,D^*}~~~ 
 \Longrightarrow~~~ m_{X_u} \ll m_{X_c}.$$
In Fig. 1 we show the double differential distribution \cite{our},
$\frac{d\Gamma}{dm_X dE_l}$.

\begin{figure}[tb]
\vspace*{-1cm}
\psfig{figure=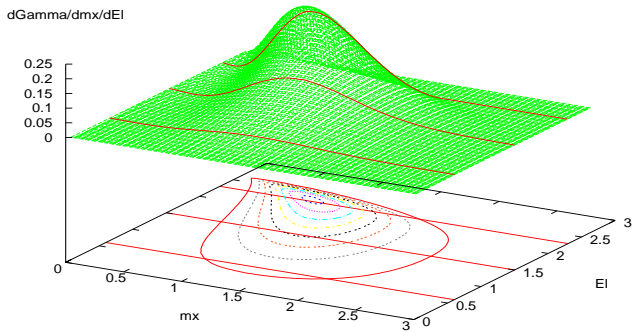,height=6cm,width=10cm,angle=-90}
\caption{3-dimension plot of $\frac{d\Gamma}{dm_X dE_l}$ in 
$B \to X_u l \nu$.}
\end{figure}

\subsection{$|V_{ts}|$ and $|V_{td}|$}

$B_d-\overline{B}_d$ and $B_s-\overline{B}_s$ mixings can give values of
$|V_{td}|$ and $|V_{ts}|$ from
\begin{eqnarray}
\Delta m_q &\equiv& m(B_q^H)-m(B_q^L) \nonumber \\
 &=& \left(\frac{G_F^2}{6 \pi^2} m_W^2 m_{B_q}\right)[f_{B_q}^2 B_{B_q}]
[\eta_{\rm QCD} {\cal F}_{\rm sd}^q] |V_{tq}^* V_{tb}|^2, 
\end{eqnarray}
where $f_{B_d}\approx (170 \sim 180)\pm 30$ MeV, 
$f_{B_s}\approx (195\sim 205)\pm 30$ MeV, 
$\frac{f_{B_s}}{f_{B_d}} = 1.15 \pm 0.05$, 
$B_{B_d}=0.75 \pm 0.15 \approx B_{B_s}$ and
$f_{B_s}\sqrt{B_{B_s}}/f_{B_d}\sqrt{B_{B_d}} = 1.14 \pm 0.06$.
The recent experimental values \cite{Vtds} are
\begin{equation}
\Delta m(B_d)=0.471 \pm 0.016~{\rm ps}^{-1},~~~
\Delta m(B_s) > 14.3~{\rm ps}^{-1}~(95\%~{\rm CL}).
\end{equation}
If we assume the Standard Model short distance (sd) interaction, 
${\cal F}_{\rm sd}^d = {\cal F}_{\rm sd}^s$, then: 
\begin{itemize}
\item $\Delta m(B_d)_{\rm exp}$ gives 
 $|V_{td}| \approx (8.1 \pm 1.8)\times 10^{-3}$.
\item Asumming  $|V_{ts}|=|V_{cb}|$ and $|V_{td}|=(0.004 \sim 0.012)$,
 we get 
$$
\left(\frac{\Delta m(B_d)}{\Delta m(B_s)}\right)_{\rm SM} =0.056 \pm 0.08,
$$
which is quite larger than the present experimental
ratio, $< 0.0341$ (95$\%$ CL).
\item Assuming the Standard Model is correct, then the present data (9) gives
$|V_{td}/V_{ts}|< 0.217$  (95$\%$ CL).
\end{itemize}
As is well known, the mass difference, $\Delta m_q$, can be easily poluted
by new physics, and so we can approach differently to find new physics,
instead of determining the Standard Model parameters, like $V_{ts},V_{td}$.
Any new physics, that has different short distance (sd) interaction structure,
or has spectator-quark depending interactions, is likely to show up 
in $\Delta m_q$.

\section{Generalization of $V_{\rm CKM}$ and Unitary Conditions}
\subsection{Generalization of $V_{\rm CKM}$}

If we are considering the possibility of non-unitary CKM matrix,
Eq. (1) can be generalized now with 4 phase angles 
(after absorbing 5 phases to quark fields),
and nine real mixing angles parametrized by $|V_{ij}|$, resulting to 13
independent parameters in total: 
\begin{equation}
V_{\rm CKM}^{\rm general} = \left(
   \begin{array}{ccc}
     |V_{ud}|   &   |V_{us}|   & |V_{ub}| e^{i\delta_{13}} \\
     |V_{cd}|   &   |V_{cs}| e^{i\delta_{22}}  & |V_{cb}|  \\
     |V_{td}| e^{i\delta_{31}}   &   |V_{ts}|  & |V_{tb}| e^{i\delta_{33}}
   \end{array}
   \right).
\end{equation}
Based on present experimental constraints on the
values of CKM matrix  elements, and starting from the approximate Wolfenstein
parameterization (4), which is already non-unitary, 
the generalized CKM matrix
can be parametrized as
\begin{equation}
V_{\rm CKM}^{\rm general}  \approx 
  \left( 
  \begin{array}{ccc}
 	  1-\frac{\lambda^2}{2}      &  \lambda     &  
     A_1\lambda^3(\rho_1-i\eta_1) \\  -\lambda &
        1-\frac{\lambda^2}{2} & A_1\lambda^2 \\
     A_1 \lambda^3 (1-\rho_2-i \eta_2) & A_2\lambda^2 & 1  
  \end{array} 
  \right) . 
\end{equation}
We note that 
\begin{itemize}
\item By assuming that $V_{tb}=1$ and $V_{cs}$ is real
as in the case of Wolfenstein parametrization, in Eq. (11) 
we now have only 5 real parameters,  $\lambda, A_1, A_2, \rho_1, \rho_2$
and 2 phase parameters, $\eta_1, \eta_2$.
Except for $V_{tb}$, the values of elements
in the right-most column or the bottom row are not constrained by the
parametrization.
\item $|V_{cb}|$ and $|V_{ts}|$ can be different because of
$A_1 \ne A_2$. This will produce consequently many new results  in analysis
of CKM constraints, such as $B_s - \overline B_s$ mixing, $B \to X_s +\gamma$
decay, $b \to s$ penguin, etc. 
\item In order to check the unitarity
  of CKM matrix, we have to use the general parametrization (10) or (11)
  instead of the already unitarized parametrization (1) or (4) in 
  extracting CKM parameters from the experimental
  observables.
\item If there exists a massive singlet down quark $b^\prime$, then the columns
of $V_{\rm CKM}^{\rm general}$  remain to be unitary, but the rows
do not. 
\end{itemize}
Usual unitary relation for the matrix (11) becomes
\begin{eqnarray}
V_{k1} V_{k3}^* &=& V_{11}V_{13}^*+V_{21}V_{23}^*+V_{31}V_{33}^* = 0~? \\
0 &\simeq& V_{13}^* -A_1 \lambda^3 +V_{31} , \nonumber \\
0 &\approx& (A_1 \lambda^3) \times 
\{ (\rho_1+i \eta_1) -1 + (1-\rho_2 -i \eta_2) \} ,
\end{eqnarray}
and now we have only 5 unknowns,  
 $$
A_1\lambda^3,~~ \rho_1,~~ \eta_1,~~ \rho_2~~~{\rm and}~~~\eta_2,
$$
or equivalently
$$
\lambda|V_{cb}|,~~ |V_{ub}|,~~ \sin\gamma,~~ |V_{td}|~~~{\rm and}~~~
\sin\beta.
$$
Those 5 unknowns are exactly the same as the 5 sufficient conditions for the
drawing a triangle uniquely -- three sides and two angles.
As is well known, the usual 3 conditions, two sides and one angle, which are
$A_1\lambda^3,~\rho,~\eta$ (or $\lambda|V_{cb}|,~|V_{ub}|,~\sin\gamma$
or equivalently  $A_1\lambda^3,~|V_{td}|,~\sin\beta $) in
unitarized CKM matrix, are only the  necessary conditions for drawing a
triangle. In order to see that Eq. (13) equals zero, we have to measure those
5 independent observables.

\subsection{(Minimum and Complete) Unitary Conditions}

As is explained, we have to measure precisely 5 observables to check if just
(most popular) one of unitary triangles is really triangle.
Eq. (13) can be written in two equations by using the usual sine and cosine
rules,
\begin{eqnarray}
{\rm (A)}~~~ \frac{|V_{ub}|}{\sin\beta} &=& \frac{|V_{td}|}{\sin\gamma}~~~~
{\rm or}~~~ \frac{|V_{td}|}{|V_{ub}|} = \frac{\sin\gamma}{\sin\beta} , \\
{\rm (B)}~~ \frac{\lambda|V_{cb}|}{|V_{ub}|} &=& \cos\gamma +
 \frac{|V_{td}|}{|V_{ub}|} \cos\beta . \nonumber
\end{eqnarray}
Here we have, instead of one sine (or cosine) rule, two equations because
the third angle $\alpha$ is {\it not} independent due to the relation
$\alpha+\beta+\gamma=\pi$. We remark here a few comments on Eq. (14):
\begin{itemize}
\item The angles $\beta,~\gamma$ need to be measured independently without
assuming unitarity of CKM matrix. Recent direct measurement of
$\beta$ at CDF still relies on the presumed unitary assumption of
$|V_{cb}|=|V_{ts}|$ in the analysis of $B \to J/\psi K_s$ decay.
If we ignore the small unitary violation in $|V_{cb}| \ne |V_{ts}|$,
CDF's direct result \cite{CDF} can be one solid ingredient for Eq. (14) as
$$
\beta_{\rm CDF} =0.79 \pm 0.43.
$$
\item Future measurements on $\frac{|V_{ub}|}{|V_{cb}|}$ at Babar and Belle,
as explained in section 2,
would be one of the most important ingredient in the test of Eq. (14B).
\item For the ratio $\frac{|V_{td}|}{|V_{ub}|}$, 
we may use the relation \cite{Aliev},
\begin{equation}
\frac{{\cal BR}(B^\pm \to \rho^\pm \nu \bar\nu)}
     {{\cal BR}(B^\pm \to \rho^0 e^\pm \nu)} =
6 \left(\frac{\alpha^2}{4 \pi^2}\right) |C_{10}^\nu|^2 \times 
\frac{|V_{td}|^2}{|V_{ub}|^2},
\end{equation}
with
$$
C_{10}^\nu \equiv \frac{X(m_t^2/m_W^2)}{\pi \sin^2\theta_W},~~~
{\rm where}~~X(x_t)~{\rm is~Inami~Lim~function}.
$$
In Eq. (15), the constant 6 comes from 3 neutrino species and from isospin
relation in form factors of $B \to \rho^\pm,~ B \to \rho^0$.
Because of complete cancelation of the hadronic form factors in the ratio
of branching fractions of those two decays, there is not any theoretical
uncertainties in Eq. (15), 
though it would be an experimental challenge to measure 
the small branching fraction, 
${\cal BR}(B \to \rho \nu \nu) \sim 4 \times 10^{-7}$.
\end{itemize}
Measuring angle $\gamma$ would be very difficult, if we donot assume
any unitarity in the analyses. However, if we assume the unitarity priorly, 
then we can calculate it from the relation (14A), and then compare the value
with the independently measured values from, 
as an example, Neubert-Rosner bound \cite{NRbound} to check if the unitarity
holds.

\subsection{Comments on the Discrepancy in recently extracted Values of
$\gamma$ }

As is well known, the discrepancy in extracted values of $\gamma$ from
CKM-fitting \cite{CKMfit}
at $\rho-\eta$ plan and from the $\chi^2$ analysis \cite{nonlep} of
non-leptonic $B$ decays has aroused hot debates over what is going on
underneath of the unitary triangle. The value of $\gamma$ has been obtained as
\begin{eqnarray}
\gamma &=& 60^0 \sim 80^0~~~~({\rm from~unitary~triangle~fitting}), \nonumber\\
       &=& 90^0 \sim 140^0~~~({\rm from~nonleptonic~analysis}). \nonumber
\end{eqnarray}
In both analyses, the unitary conditions have been extensively assumed.
If we believe both analyses are correctly performed and
the theoretical assumptions used (including the factorization assumption) are
correct, one of the most plausible answers would be ``non-unitarity of CKM
matrix". An easy answer from our previous argument, explained in the second item
after Eq. (11), is that $|V_{cb}| \ne |V_{ts}|$.
If those two elements are not equal, then:
\begin{itemize}
\item We can not simply add the constraint from $B_s - \overline B_s$ mixing
result on $\rho-\eta$ plane to get the allowed smaller circular region of
$|V_{td}|$. 
\item Adding of two circular regions (from measurements $|V_{ub}|, ~|V_{td}|$) 
does not reduce to the overlapped small region. 
Instead, we will have the summed large region
in the unitary plane. Therefore, without the direct measurement on $\gamma$ we 
cannot decide which value is correct.
\end{itemize}
Very soon, we will have a flood of experimental results on CKM elements
from many experiments, asymmetric and symmetric $e^+e^-$ colliders, 
and hadronic machines.
Without having correct (theoretical) stratage of coping the data, we will
be easily fooled. We argue the present discrepancy on $\gamma$ is just
an early example.

\section*{Acknowledgments}
CSK was supported 
in part by  Grant No. 1999-2-111-002-5 from the Interdisciplinary 
research program of the KOSEF,
in part by the BSRI Program of MOE, Project No.
99-015-DI0032, 
and in part by the KRF Sughak-research program, 
Project No. 1997-011-D00015. HY was supported by the Department of Energy
Grant DE-FG02-91ER40654.


\section*{References}
\begin{thebibliography}{99}

\bibitem{CKM} N. Cabibbo, \Journal{\PRL}{10}{531}{1963};
   M. Kobayashi, T. Maskawa, \Journal{\PROG}{49}{652}{1973}.
\bibitem{PDG} Particle Date Group, \Journal{\EPJC}{3}{1}{1998}.
\bibitem{Wolf} L. Wofenstein, \Journal{\PRL}{51}{1945}{1983}.
\bibitem{Neubert} M. Neubert, hep-ph/9801269.
\bibitem{CLEO} P. Drell, hep-ex/9711020.
\bibitem{Vqb} M. Luke {\it et al.,} \Journal{\PLB}{321}{88}{1994};
M. Shifman {\it et al.,} \Journal{\PRD}{51}{2217}{1995}; 
P. Ball {\it et al.,} \Journal{\PRD}{52}{3929}{1995}.
\bibitem{CLEO2} CLEO Collab., \Journal{\PRD}{61}{052001}{2000}. 
\bibitem{Vub} N. Uraltzev, \Journal{\IJMPA}{11}{515}{1996};    
   C.S. Kim, \Journal{\PLB}{414}{347}{1997}.
\bibitem{Barger} V. Barger, C.S. Kim, R.J.N. Phillips, 
   \Journal{\PLB}{251}{629}{1990}.
\bibitem{our} K.K. Jeong, C.S. Kim, Y.G. Kim, work in progress.
\bibitem{Vtds} $|V_{ts}|,|V_{td}|$ from Lepton-Photon Symposium (1999).
\bibitem{CDF} CDF Collab., hep-ex/9909003.
\bibitem{Aliev} T. Aliev, C.S. Kim, \Journal{\PRD}{58}{013003}{1998}.
\bibitem{NRbound} M. Neubert, J.R. Rosner, \Journal{\PRL}{81}{5076}{1998}.
\bibitem{CKMfit} A. Stocchi, talk given in this Conference.
\bibitem{nonlep} N.G. Deshpande, X. He, W. Hou,
   S. Pakvasa,  \Journal{\PRL}{82}{2240}{1999}.
\end{thebibliography}
\end{document}